\documentclass[twocolumn,showpacs,prl,aps,amssymb]{revtex4}

\usepackage{graphicx}% Include figure files
\usepackage{bm}% bold math

\begin{document}

\title{Exact Duality Relations in Correlated Electron Systems}

\author{Tsutomu Momoi$^{1,2}$ and Toshiya Hikihara$^1$}
\affiliation{$^1$Condensed-Matter Theory Laboratory, RIKEN,
%(The Institute of Physical and Chemical Research),
Wako, Saitama 351-0198, Japan\\
$^2$Institute of Physics, University of Tsukuba, Tsukuba, Ibaraki
305-8571, Japan}
%\date{\hspace*{5cm}}
\date{\today}

\begin{abstract}
Using gauge transformations on electron bond operators, we derive
exact duality relations between various order parameters for
correlated electron systems. Applying these transformations, we
find two duality relations in the generalized two-leg Hubbard
ladder at arbitrary filling. The relations show that
unconventional density-wave orders such as staggered flux or
circulating spin current are dual to conventional density-wave
orders and there are direct mappings between dual phases. Several
exact results on the phase diagram are also concluded.

\end{abstract}

\maketitle

%background, motivation
Unconventional density-wave orders such as staggered flux [which
is equivalently called as $d$-density wave (DDW) or orbital
antiferromagnets] and circulating spin current were first proposed
in the context of excitonic insulators\cite{HalperinR} and later
discussed in high-$T_c$
superconductors\cite{AffleckM,Schultz,NersesyanV}, but the
appearance of these orders was not established at that time.
Recently, several experimental results have led to a resurgence of
interest in the possibility of the exotic orders. A circulating
spin-current state\cite{IkedaO} and DDW state\cite{ChandraCMT}
were proposed as an origin of hidden order in the heavy-fermion
compound URu$_2$Si$_2$. A DDW state has also been discussed to
appear in underdoped region of high-$T_c$
superconductors\cite{ChakravartyLMN}, where a pseudogap was
observed\cite{TimuskS}, and in the quasi-two-dimensional organic
conductor $\alpha$-(ET)$_2$KHg(SCN)$_4$\cite{DoraMV}.

From a theoretical viewpoint, the possibility of the
unconventionally ordered phases in microscopic models of
correlated electrons has been the focus of interest. In
particular, the two-leg ladder system is attracting attention as a
minimal model for showing the exotic orders. Though the two-leg
Hubbard and $t$-$J$ ladders do not show any
order\cite{DagottoR,MarstonFS,Schollwoeck}, both strong coupling
and weak coupling analyses revealed that the generalized two-leg
Hubbard ladder including nearest-neighbor interactions exhibits
various ordered phases at
half-filling\cite{LinBF,FjaerestadM,TsuchiizuF,WuLiuF} and doped
cases\cite{WuLiuF,OrignacC}. Large-scale numerical calculations
also reported the DDW phase\cite{MarstonFS} at half-filling and
charge-density-wave (CDW)\cite{VojtaHN} and DDW\cite{Schollwoeck}
phases for less than half-filling.
%Large-scale numerical studies are being performed searching for
%various possible exotic phases.

%summary
\begin{figure}[b]
  \centering
  \includegraphics[width=85mm]{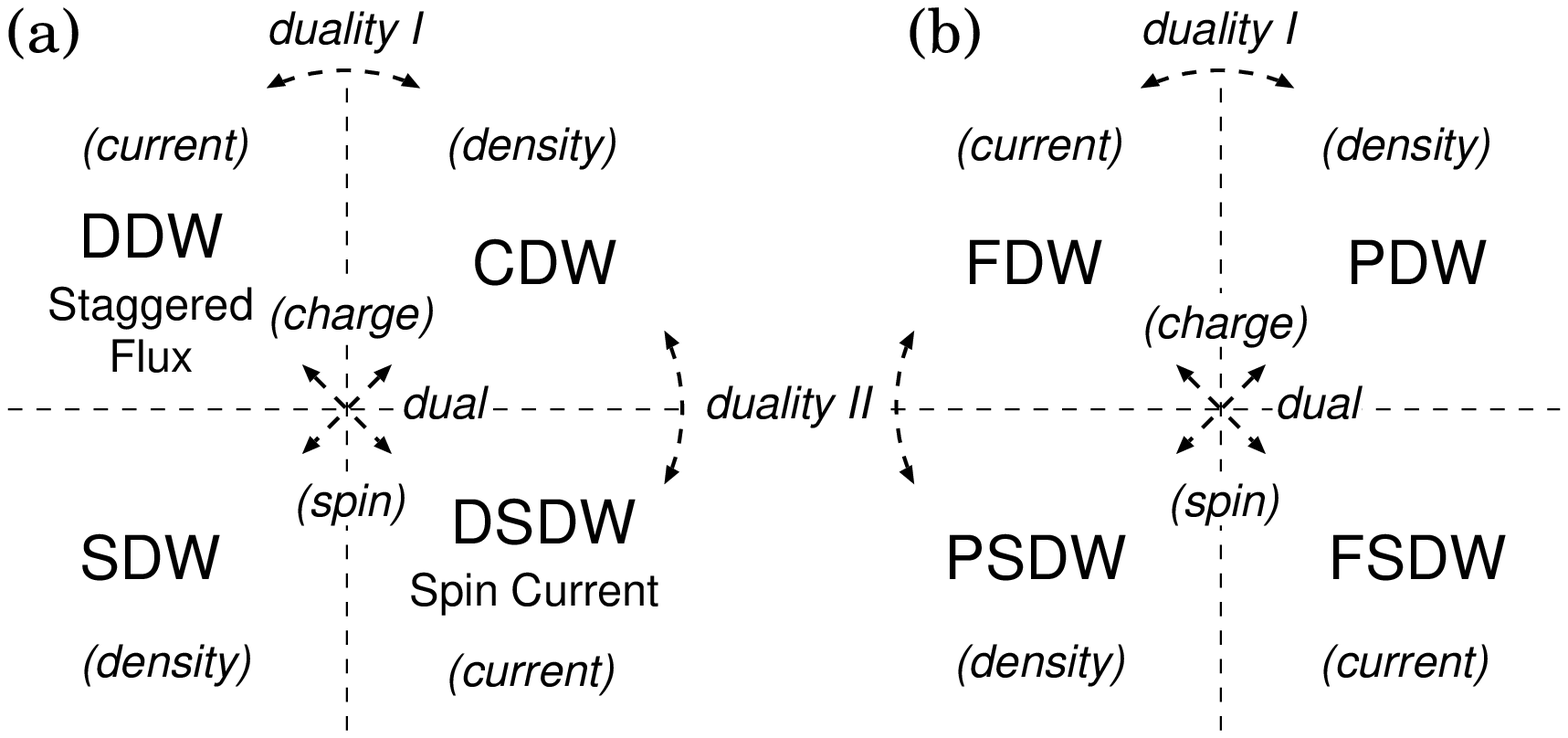}\medskip\\
  \includegraphics[width=35mm]{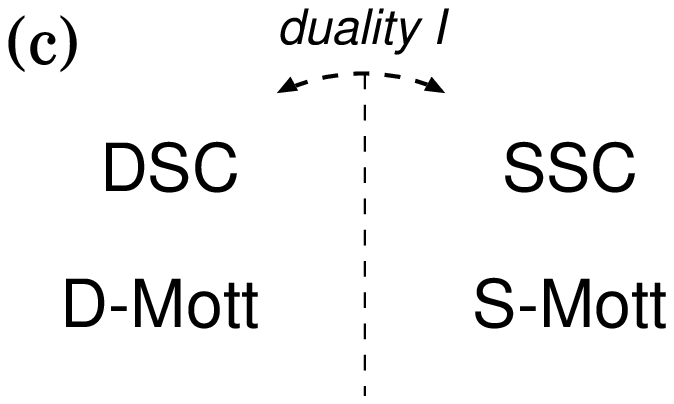}
  \caption{Duality relations (I) and (II) between various states;
  (a) CDW, DDW, SDW, and DSDW states,
  (b) PDW, FDW, PSDW, and FSDW states, and
  (c) $s$- and $d$-wave Mott-insulating and superconducting
  states.}\label{fig:duality}
\end{figure}
In this Letter, introducing gauge transformations on electron bond
operators, we derive two exact duality relations between various
exotic phases in correlated electron systems. One duality relates
conventional density waves to unconventional density waves (or
currents) and the other relates charge-density degrees of freedom
to spin-current ones. These transformations also show duality
relations between $s$- and $d$-wave superconductivity (SSC and
DSC), as well as between $s$- and $d$-wave Mott-insulating (S-Mott
and D-Mott) states. Applying these transformations, we show two
exact duality relations in the generalized two-leg Hubbard ladder
(or similarly the Hubbard model with doubly degenerate orbitals).
The duality relations among various phases are summarized in Fig.\
\ref{fig:duality}, where we use density-wave
terminology\cite{TsuchiizuF}. Starting from previously established
phases, we can reveal the appearance of various exotic phases with
and without spin symmetry breaking. The transformations give
parameter mapping between dual phases and fix boundaries between
them on self-dual lines.
%When
%one finds an ordered phase, one can naturally conclude appearance
%of dual ordered phases in dual space.
These exact relations enable us to clarify the nature of the phase
diagram and find the parameter space of unconventionally ordered
phases.

%order parameter
Before discussing the duality transformations, we list operators
and order parameters considered in this work. We use ladder
geometry in the following for simplicity, but one can extend to
other lattices adopting appropriate bonds. Regarding charge
degrees of freedom, we consider CDW, staggered-flux [or
equivalently DDW], staggered dimer [or $p$-density-wave (PDW)],
and diagonal current [or $f$-density-wave (FDW)] operators on
rungs (or plaquettes) given by
\begin{eqnarray}
{\cal O}_{\rm CDW} (j) &=& \frac{1}{2} \sum_\sigma
(n_{j,1,\sigma}-n_{j,2,\sigma}), \label{eq:O_CDW}\\
{\cal O}_{\rm DDW} (j) &=& \frac{i}{2} \sum_\sigma
(c_{j,1,\sigma}^\dagger
c_{j,2,\sigma} - {\rm H.c.}), \label{eq:O_DDW}\\
{\cal O}_{\rm PDW} (j) &=& \frac{1}{4} \sum_\sigma
[(c_{j+1,1,\sigma}^\dagger c_{j,1,\sigma} -
c_{j+1,2,\sigma}^\dagger c_{j,2,\sigma})\nonumber\\
& &~~~~~ + {\rm H.c.} ], \label{eq:O_PDW}\\
{\cal O}_{\rm FDW} (j) &=& \frac{i}{4} \sum_\sigma
[(c_{j+1,1,\sigma}^\dagger c_{j,2,\sigma} -
c_{j+1,2,\sigma}^\dagger c_{j,1,\sigma}) \nonumber\\
& &~~~~~ -{\rm H.c.}], \label{eq:O_FDW}
\end{eqnarray}
respectively, where $c^\dagger_{j,\mu,\sigma}$
($c_{j,\mu,\sigma}$) ($\mu=1,2$ and $\sigma=\uparrow,\downarrow$)
denotes an electron creation (annihilation) operator on the
$\mu$th site of the $j$th rung and $n_{j,\mu,\sigma} =
c^\dagger_{j,\mu,\sigma}c_{j,\mu,\sigma}$. We note that the CDW
and PDW orders are kinds of density waves while the DDW and FDW
orders finite local currents with broken time-reversal symmetry.
%Order parameters on the ladder are given by
%\[
%\frac{1}{L}\sum_j {\cal O}_{A}(j)\exp(iqj)
%=\frac{1}{2L}\sum_{{\bf
%k},\sigma}f_A({\bf k}) c^\dagger_\sigma({\bf k}) c_\sigma({\bf
%+Q}),
%\]
%where $L$ is the number of rungs,
%${\bf k}=(k_\parallel,k_\perp)$,
%${\bf Q}=(q,\pi)$, and $c_\sigma({\bf k})=\sum_j e^{ik_\parallel
%j} (c_{j,1,\sigma}+e^{ik_\perp}c_{j,2,\sigma})/\sqrt{L}$. The form
%factor $f_A({\bf k})$ is defined as $f_{\rm CDW}({\bf k})=1$,
%$f_{\rm DDW}({\bf k})=\cos k_\perp$, $f_{\rm PDW}({\bf k})=\sin
%k_\parallel$, and $f_{\rm FDW}({\bf k})=\sin k_\parallel \cos
%k_\perp$.
In the similar way, we consider operators in the spin sector.
Inserting the Pauli matrix $\sigma^z_{\sigma,\sigma}$ into the
right-hand sides of Eqs.\ (\ref{eq:O_CDW})-(\ref{eq:O_FDW}), one
defines spin-density wave (SDW) operator ${\cal O}_{\rm SDW}~[=
\frac{1}{2} \sum_\sigma \sigma^z_{\sigma,\sigma}
(n_{j,1,\sigma}-n_{j,2,\sigma})]$, circulating spin current [or
$d$-spin-density wave (DSDW)] operator ${\cal O}_{\rm DSDW}~[=
\frac{i}{2} \sum_\sigma \sigma^z_{\sigma,\sigma}
(c_{j,1,\sigma}^\dagger c_{j,2,\sigma} - {\rm H.c.})]$,
$p$-spin-density-wave (PSDW) operator ${\cal O}_{\rm PSDW}$, and
$f$-spin-density wave (FSDW) operator ${\cal O}_{\rm FSDW}$ as the
counterparts of the CDW, DDW, PDW, and FDW operators,
respectively. Order parameters on the ladder are given by
%$O_A(q)=L^{-1} \sum_j {\cal O}_{A}(j)\exp(iqj)$,
\[
O_A(q)=L^{-1} \sum_j {\cal O}_{A}(j)\exp(iqj),
\]
%=\frac{1}{2L}\sum_{{\bf
%k},\sigma}f_A({\bf k}) c^\dagger_\sigma({\bf k}) c_\sigma({\bf
%+Q})
where $L$ is the number of rungs.
%Corresponding order
%parameters for the spin sector are given by
%\[
%\frac{1}{2L}\sum_{{\bf k}}\sum_{\sigma=\uparrow,\downarrow}
%f_A({\bf k}) c^\dagger_\sigma({\bf k}) \sigma^z_{\sigma \sigma}
%c_{\sigma}({\bf k+Q}),
%\]
%where $f_A$ is the same as that for charge counterparts.
Furthermore, we consider the $d$- and $s$-wave pairing operators,
\begin{eqnarray}
{\cal O}_{\rm DSC} (j) &=& \frac{1}{\sqrt{2}}
(c_{j,1,\uparrow}c_{j,2,\downarrow} - c_{j,1,\downarrow}
c_{j,2,\uparrow}),
\\
{\cal O}_{\rm SSC} (j) &=& \frac{1}{\sqrt{2}}
(c_{j,1,\uparrow}c_{j,1,\downarrow} + c_{j,2,\uparrow}
c_{j,2,\downarrow}),
\end{eqnarray}
which characterize the DSC and SSC states. We also consider the D-
and S-Mott states at half-filling,
\begin{eqnarray}
|\mbox{D-Mott} \rangle = \prod_j {\cal O}^\dagger_{\rm DSC} (j)
|0\rangle,~~~~~ |\mbox{S-Mott} \rangle = \prod_j {\cal
O}^\dagger_{\rm SSC} (j) |0\rangle,\nonumber
%|\mbox{D-Mott} \rangle &=& \prod_j {\cal O}^\dagger_{\rm DSC} (j)
%|0\rangle, \\
%|\mbox{S-Mott} \rangle &=& \prod_j {\cal O}^\dagger_{\rm SSC} (j)
%|0\rangle,
\end{eqnarray}
where $|0 \rangle$ is the vacuum of the electron operator $c_{j,\mu,\sigma}$.

Let us move to the introduction of duality transformations on
electron operators. They are given by gauge transformations on the
bonding and antibonding operators $d_{j,\pm,\sigma} =
(c_{j,1,\sigma} \pm c_{j,2,\sigma})/\sqrt{2}$. We consider two
duality transformations and their combination in the following.
%We shall show their
%application on the generalized two-leg Hubbard ladder later.

%duality I
{\it Duality I: density and current.}--- Let us consider a gauge
transformation of antibonding operators given by the unitary
operator $U_I(\theta)=\prod_{j,\sigma} \exp ( -i \theta
d^\dagger_{j,-,\sigma} d_{j,-,\sigma} )$. When $\theta=\pi/2$,
this operator yields the duality transformation (I) as
\begin{equation}
\tilde{d}_{j,+,\sigma} = d_{j,+,\sigma},~~~~~~~
\tilde{d}_{j,-,\sigma} = i d_{j,-,\sigma} \label{eq:dual1}
\end{equation}
for $\sigma=\uparrow,\downarrow$. In terms of the electron
operators $c_{j,\mu,\sigma}$, this transformation is written as
$\tilde{c}_{j,1,\sigma} = \frac{1+i}{2} c_{j,1,\sigma} +
\frac{1-i}{2} c_{j,2,\sigma}$ and $\tilde{c}_{j,2,\sigma} =
\frac{1-i}{2} c_{j,1,\sigma} + \frac{1+i}{2} c_{j,2,\sigma}$.
%This nonuniform gauge transformation changes current degrees of
%freedom. Indeed, considering the order parameters,
One finds that the duality transformation relates the density-wave
operators to the current ones as follows:
\begin{eqnarray}
\tilde{\cal O}_{\rm CDW} = - {\cal O}_{\rm DDW}, &~~~~~~&
\tilde{\cal O}_{\rm DDW} = {\cal O}_{\rm CDW},\nonumber\\
\tilde{\cal O}_{\rm SDW} = - {\cal O}_{\rm DSDW},&~~~~~~&
\tilde{\cal O}_{\rm DSDW}={\cal O}_{\rm SDW},\nonumber\\
\tilde{\cal O}_{\rm PDW} = - {\cal O}_{\rm FDW}, &~~~~~~&
\tilde{\cal O}_{\rm FDW} = {\cal O}_{\rm PDW}, \label{eq:dualityI}\\
\tilde{\cal O}_{\rm PSDW} = - {\cal O}_{\rm FSDW},&~~~~~~&
\tilde{\cal O}_{\rm FSDW} = {\cal O}_{\rm PSDW}.\nonumber
\end{eqnarray}
Thus, current phases, i.e., DDW, DSDW, FDW, and FSDW phases, turn
out to be dual to density-wave phases, i.e., CDW, SDW, PDW, and
PSDW phases, respectively. In other words, this transformation
changes the bond parity and hence $s$- ($p$-) wave orders are
converted to $d$- ($f$-) wave ones. The pairing operators and the
Mott states are transformed in the similar way,
\begin{eqnarray}
\tilde{\cal O}_{\rm DSC} &=& {\cal O}_{\rm SSC},~~~~~~~~~
\tilde{\cal O}_{\rm SSC} = {\cal O}_{\rm DSC}, \nonumber \\
\widetilde{|\mbox{D-Mott}\rangle} &=& |\mbox{S-Mott}\rangle,~~~~~
\widetilde{|\mbox{S-Mott}\rangle} = |\mbox{D-Mott}\rangle.
\nonumber
\end{eqnarray}
We note that the unitary operator $U_I (\theta)$ gives a
continuous transformation between dual operators, e.g., $U_I
(\theta){\cal O}_{\rm CDW}U_I (\theta)^{-1}= {\cal O}_{\rm
CDW}\cos \theta
 - {\cal O}_{\rm DDW}\sin \theta$.

%duality II
{\it Duality II: density and spin current.}--- Next, we consider a
gauge transformation given by the unitary operator
$U_{II}(\theta)=\prod_{j} \exp [ -i \theta
(d^\dagger_{j,+,\uparrow} d_{j,+,\uparrow} +
d^\dagger_{j,-,\downarrow} d_{j,-,\downarrow}) ]$. This operator
with $\theta=\pi/2$ gives the duality transformation (II) as
\begin{eqnarray}
\begin{array}{lc}
  \bar{d}_{j,+,\uparrow} = i d_{j,+,\uparrow},~~~~~~~
  & \bar{d}_{j,+,\downarrow} = d_{j,+,\downarrow}, \\
  \bar{d}_{j,-,\uparrow} = d_{j,-,\uparrow},
  & \bar{d}_{j,-,\downarrow} = i d_{j,-,\downarrow}.\\
\end{array}
\label{eq:dual2}
\end{eqnarray}
%One can see that this spin-dependent gauge transformation changes
%spin current degrees of freedom. Indeed,
This transformation converts the operators as
\begin{eqnarray}
\bar{\cal O}_{\rm CDW} = {\cal O}_{\rm DSDW}, &~~~~~~&
\bar{\cal O}_{\rm DSDW}=-{\cal O}_{\rm CDW},\nonumber\\
\bar{\cal O}_{\rm DDW} = -{\cal O}_{\rm SDW}, &~~~~~~&
\bar{\cal O}_{\rm SDW}={\cal O}_{\rm DDW},\nonumber\\
\bar{\cal O}_{\rm PDW} = {\cal O}_{\rm FSDW}, &~~~~~~&
\bar{\cal O}_{\rm FSDW}=-{\cal O}_{\rm PDW},\label{eq:dualityII}\\
\bar{\cal O}_{\rm FDW} = -{\cal O}_{\rm PSDW},&~~~~~~& \bar{\cal
O}_{\rm PSDW} = {\cal O}_{\rm FDW}.\nonumber
%\label{eq:dualII_op4}
\end{eqnarray}
One can see that density waves of charges are transformed into
spin currents while density waves of spins into charge currents.
This transformation thus exchanges density and current as well as
spin and charge degrees of freedom. Similarly to $U_{I}(\theta)$,
the unitary operator $U_{II} (\theta)$ gives a continuous
transformation between dual operators such as $U_{II}
(\theta){\cal O}_{\rm CDW}U_{II} (\theta)^{-1}= {\cal O}_{\rm
CDW}\cos \theta + {\cal O}_{\rm DSDW}\sin \theta$. We note that
under this transformation the pairing operators and Mott states
are invariant except for phase factors.

%spin-charge
{\it Duality between spin and charge.}--- Combining the duality
transformations (\ref{eq:dual1}) and (\ref{eq:dual2}), we obtain
the transformation
\begin{eqnarray}
\begin{array}{lc}
  \hat{d}_{j,+,\uparrow} = d_{j,+,\uparrow},~~~~~~~
  & \hat{d}_{j,+,\downarrow} = -i d_{j,+,\downarrow}, \\
  \hat{d}_{j,-,\uparrow} = d_{j,-,\uparrow},
  & \hat{d}_{j,-,\downarrow} = i d_{j,-,\downarrow}.\\
\end{array}
\label{eq:dual_comb}
\end{eqnarray}
%This transforms order parameters as
%\begin{eqnarray}
%\tilde{\cal O}_{\rm CDW} &=& {\cal O}_{\rm SDW},~~~~~~
%\tilde{\cal O}_{\rm SDW}=-{\cal O}_{\rm CDW},\\
%\tilde{\cal J} &=& {\cal J}_{\rm s},~~~~~~~~
%\tilde{\cal J}_{\rm s}=-{\cal J},\\
%\tilde{\cal O}_{\rm PDW} &=& {\cal O}_{\rm PSDW},~~~~~~
%\tilde{\cal O}_{\rm PSDW}=- {\cal O}_{\rm PDW},\\
%\tilde{\cal O}_{\rm FDW} &=& {\cal O}_{\rm FSDW},~~~~~~
%\tilde{\cal O}_{\rm FSDW}=-{\cal O}_{\rm FDW}.
%\end{eqnarray}
From the duality relations (\ref{eq:dualityI}) and
(\ref{eq:dualityII}), it follows that this transformation
exchanges spin and charge degrees of freedom, e.g., density-wave
operators
%\begin{equation}
%O(q)=\frac{1}{2L}\sum_{{\bf k},\sigma}f({\bf k})
%c^\dagger_\sigma({\bf k}) c_\sigma({\bf k+Q})
%\end{equation}
are transformed to spin-density-wave operators.
%as $\tilde{O}_A(q)=O_{SA}(q)$.
%\begin{equation}
%\tilde{O}(q)=O_{S}(q)=\frac{1}{2L}\sum_{{\bf
%k}}\sum_{a,b=\uparrow,\downarrow} f({\bf k}) c^\dagger_a({\bf k})
%\sigma^z_{ab} c_b({\bf k+Q}),
%\end{equation}
%where ${\bf k}=(k_\parallel,k_\perp)$ and ${\bf Q}=(q,\pi)$.

%Application to extended Hubbard model
To elucidate efficiency of these transformations, we apply them to
the generalized two-leg Hubbard ladder with intra-rung couplings.
The Hamiltonian is defined by
\begin{eqnarray}
H &=& -\sum_{j,\sigma} \left( t_\parallel
%\sum_{\mu = 1,2} (c_{j,\mu,\sigma}^\dagger c_{j+1,\mu,\sigma} + {\rm
%H.c.})}
%nonumber\\
%&+& t_\perp (c_{j,1,\sigma}^\dagger c_{j,2,\sigma} + {\rm
%.c.}) ]\nonumber\\
%
\sum_{\mu = 1,2} c_{j,\mu,\sigma}^\dagger c_{j+1,\mu,\sigma} + t_\perp
c_{j,1,\sigma}^\dagger c_{j,2,\sigma} + {\rm H.c.} \right)
\nonumber\\
& &+ \sum_j \Biggl[U  \sum_{\mu=1,2} n_{j,\mu,\uparrow}
n_{j,\mu,\downarrow} + V_\perp n_{j,1} n_{j,2} \Biggr.
\nonumber\\
& &+ J_\perp (S^x_{j,1} S^x_{j,2}+S^y_{j,1} S^y_{j,2}) + J^z_\perp
S^z_{j,1} S^z_{j,2}
\nonumber\\
& &+ \Biggl. t_{\rm pair} (c_{j,1,\uparrow}^\dagger
c_{j,1,\downarrow}^\dagger c_{j,2,\downarrow} c_{j,2,\uparrow} +
{\rm H.c.})\Biggr], \label{eq:Ham}
\end{eqnarray}
where $n_{j,\mu}=\sum_\sigma n_{j,\mu,\sigma}$ and ${\bm
S}_{j,\mu} = \frac{1}{2}
\sum_{\sigma,\sigma^\prime}c^\dagger_{j,\mu,\sigma} {\bm
\sigma}_{\sigma,\sigma'} c_{j,\mu,\sigma'}$. When $t_\perp=0$,
this Hamiltonian includes the Hubbard model with doubly degenerate
orbitals\cite{KugelK}. We first apply the transformation (I) [Eq.\
(\ref{eq:dual1})] to the Hamiltonian. It can be shown that this
transformation maps the model onto the same Hubbard model with
different coupling parameters. The total charge density, the total
magnetization, and the kinetic energy terms are invariant under
the transformation. However, the coupling terms are mixed by the
transformation. Here, it is useful to rewrite the couplings in
terms of bonding and antibonding operators as
%\begin{widetext}
%\begin{eqnarray}
%& & \left(\frac{1}{4}V_\perp+\frac{1}{16}J_\perp\right)
%\sum_{\alpha,\sigma} d_{j,\alpha,\sigma}^\dagger
%d_{j,\alpha,\sigma}
%+\frac{1}{2}\left(U - V_\perp + \frac{3}{4}J_\perp + t_{\rm
%pair}\right) (d_{j,+,\uparrow}^\dagger d_{j,-,\uparrow}
%d_{j,+,\downarrow}^\dagger d_{j,-,\downarrow} + {\rm H.c.})
%\nonumber\\
%&+&\frac{1}{2}\left(U - V_\perp -\frac{1}{4}J_\perp - t_{\rm
%pair}\right) (d_{j,+,\uparrow}^\dagger d_{j,-,\uparrow}
%d_{j,-,\downarrow}^\dagger d_{j,+,\downarrow} + {\rm H.c.})
%+\frac{1}{2}\left(U + V_\perp -\frac{3}{4}J_\perp + t_{\rm
%pair}\right) \sum_\alpha d_{j,\alpha,\uparrow}^\dagger
%d_{j,\alpha,\uparrow} d_{j,\alpha,\downarrow}^\dagger
%d_{j,\alpha,\downarrow}
%\nonumber\\
%&+&\frac{1}{2}\left(U + V_\perp +\frac{1}{4}J_\perp - t_{\rm
%pair}\right) \sum_\alpha d_{j,\alpha,\uparrow}^\dagger
%d_{j,\alpha,\uparrow} d_{j,-\alpha,\downarrow}^\dagger
%d_{j,-\alpha,\downarrow}
%, \label{eq:H_coupling}
%\end{eqnarray}
%\end{widetext}
\begin{eqnarray}
& & A (d_{j,+,\uparrow}^\dagger d_{j,-,\uparrow}
d_{j,+,\downarrow}^\dagger d_{j,-,\downarrow} + {\rm H.c.})
\nonumber\\
&+& B (d_{j,+,\uparrow}^\dagger d_{j,-,\uparrow}
d_{j,-,\downarrow}^\dagger d_{j,+,\downarrow} + {\rm H.c.})
%\nonumber\\
%&+&
+ C \sum_{\sigma} n^{(d)}_{j,+,\sigma} n^{(d)}_{j,-,\sigma} \nonumber\\
&+& D \sum_{\alpha=\pm} n^{(d)}_{j,\alpha,\uparrow}
n^{(d)}_{j,\alpha,\downarrow}
%\nonumber\\
%&+&
+ E \sum_{\alpha=\pm} n^{(d)}_{j,\alpha,\uparrow}
n^{(d)}_{j,-\alpha,\downarrow}, \label{eq:H_coupling}
%&+& C\sum_{\sigma} d_{j,+,\sigma}^\dagger d_{j,+,\sigma}
%d_{j,-,\sigma}^\dagger d_{j,-,\sigma} \nonumber\\
%&+& D \sum_{\alpha=\pm} d_{j,\alpha,\uparrow}^\dagger
%d_{j,\alpha,\uparrow} d_{j,\alpha,\downarrow}^\dagger
%d_{j,\alpha,\downarrow} \nonumber\\
%&+& E \sum_\alpha d_{j,\alpha,\uparrow}^\dagger
%d_{j,\alpha,\uparrow} d_{j,-\alpha,\downarrow}^\dagger
%d_{j,-\alpha,\downarrow}, \label{eq:H_coupling}
\end{eqnarray}
where $n^{(d)}_{j,\alpha,\sigma}=d_{j,\alpha,\sigma}^\dagger
d_{j,\alpha,\sigma}$,
%$\alpha=+$ or $-$,
and
\begin{eqnarray}
%A &=& \frac{1}{2}\left[U - V_\perp + \frac{1}{4}(2J_\perp +
%J^z_\perp) + t_{\rm pair}\right], \nonumber\\
%B &=& \frac{1}{2}\left[U - V_\perp -\frac{1}{4}(2J_\perp -
%J^z_\perp) - t_{\rm pair}\right], \nonumber\\
%C &=& V_\perp+\frac{1}{4}J^z_\perp, \nonumber\\
%D &=& \frac{1}{2}\left[U + V_\perp -\frac{1}{4}(2J_\perp +
%J^z_\perp) + t_{\rm pair}\right], \nonumber\\
%E &=& \frac{1}{2}\left[U + V_\perp +\frac{1}{4}(2J_\perp -
%J^z_\perp) - t_{\rm pair}\right].\nonumber
A &=& (U - V_\perp + t_{\rm pair})/2 + (2J_\perp +
J^z_\perp)/8, \nonumber\\
B &=& (U - V_\perp - t_{\rm pair})/2 -(2J_\perp -
J^z_\perp)/8, \nonumber\\
C &=& V_\perp+J^z_\perp/4, \nonumber\\
D &=& (U + V_\perp + t_{\rm pair})/2-(2J_\perp +
J^z_\perp)/8, \nonumber\\
E &=& (U + V_\perp - t_{\rm pair})/2+(2J_\perp -
J^z_\perp)/8.\nonumber
\end{eqnarray}
%$A = \frac{1}{2}\left[U - V_\perp + \frac{1}{4}(2J_\perp +
%J^z_\perp) + t_{\rm pair}\right]$, $B = \frac{1}{2}\left[U -
%V_\perp -\frac{1}{4}(2J_\perp - J^z_\perp) - t_{\rm pair}\right]$,
%$C = V_\perp+\frac{1}{4}J^z_\perp$, $D = \frac{1}{2}\left[U +
%V_\perp -\frac{1}{4}(2J_\perp + J^z_\perp) + t_{\rm pair}\right]$,
%$E = \frac{1}{2}\left[U + V_\perp +\frac{1}{4}(2J_\perp -
%J^z_\perp) - t_{\rm pair}\right]$.
%\end{widetext}
It is easily found in Eq.\ (\ref{eq:H_coupling}) that the
transformation (\ref{eq:dual1}) changes only the sign of the
$A$-term, but keeps the rest of terms invariant. This leads to an
exact duality relation in the two-leg Hubbard ladder. The model
with a parameter $A$ is dual to the model with $-A$ and the model
is self-dual in the space $A=0$, i.e.,
\begin{equation}
U - V_\perp + t_{\rm pair}+ (2J_\perp+J^z_\perp)/4 = 0.
\label{eq:self-dual}
\end{equation}
The mapping of the original coupling parameters is as follows:
%\begin{eqnarray}
%\tilde{U} &=& \frac{1}{2} U + \frac{1}{2} V_\perp
%  - \frac{3}{8} J_\perp - \frac{1}{2} t_{\rm pair},
%\nonumber \\
%\tilde{V_\perp} &=& \frac{1}{4} U + \frac{3}{4} V_\perp
%  + \frac{3}{16} J_\perp + \frac{1}{4} t_{\rm pair},
%\nonumber \\
%\tilde{J_\perp} &=& - U + V_\perp + \frac{1}{4} J_\perp - t_{\rm
%pair},
%\nonumber \\
%\tilde{t}_{\rm pair} &=& - \frac{1}{2} U + \frac{1}{2} V_\perp
%  - \frac{3}{8} J_\perp + \frac{1}{2} t_{\rm pair}.
%\nonumber
%\end{eqnarray}
\begin{eqnarray}
\tilde{U} &=& \frac{1}{2} (U + V_\perp - t_{\rm pair}) -
\frac{1}{8} (2 J_\perp + J_\perp^z) ,
\nonumber \\
\tilde{V}_\perp &=& \frac{1}{4} (U + 3 V_\perp + t_{\rm pair}) +
\frac{1}{16} (2 J_\perp + J_\perp^z),
\nonumber \\
\tilde{J}_\perp &=& - U + V_\perp - t_{\rm pair}+ \frac{1}{4}
(2J_\perp - J_\perp^z) ,
\label{eq:dual_param} \\
\tilde{J}_\perp^z &=& - U + V_\perp  - t_{\rm pair} - \frac{1}{4}
(2 J_\perp - 3 J_\perp^z),\nonumber\\
\tilde{t}_{\rm pair} &=& \frac{1}{2} (- U + V_\perp + t_{\rm
pair}) - \frac{1}{8} (2 J_\perp + J_\perp^z). \nonumber
\end{eqnarray}
Note that spin isotropy ($J_\perp=J_\perp^z$) is conserved through
this parameter mapping as $\tilde{J}_\perp = \tilde{J}_\perp^z$.
From the mapping (\ref{eq:dual_param}) and the duality relation
(\ref{eq:dualityI}), one can conclude that if a density-wave
order, e.g., CDW or PDW order, appears in a certain parameter
region, a dual current order, i.e., DDW or FDW order,
respectively, exists in a corresponding dual region of the
parameter space. Because of this duality relation, all phase
boundaries must be symmetric with respect to the self-dual space.
Indeed, the transition line between the CDW and DDW phases at
half-filling derived in the weak- and strong-coupling
limits\cite{LinBF,FjaerestadM,TsuchiizuF} coincides with the
self-dual line (\ref{eq:self-dual}).
%which can be quite naturally understood from the duality relation.
We stress that our result holds in general cases, regardless of
the coupling strength, filling, and system size.

%spin anisotropic extended Hubbard model
Next we discuss the duality transformation (II) [Eq.\
(\ref{eq:dual2})],
%we consider anisotropic spin coupling case,
%because the duality (II) relates density wave with spin symmetry
%breaking, which can occur only in spin anisotropic models in one
%dimension.
which gives another parameter mapping. Similarly to the
transformation (I), the total charge density, the total
magnetization, and the kinetic energy terms are invariant under
the transformation (\ref{eq:dual2}), but the coupling terms are
mixed up. From Eq.\ (\ref{eq:H_coupling}), one can see that the
transformation (\ref{eq:dual2}) changes only the sign of the
$B$-term. We thus find that the present model has another duality:
The model with a parameter $B$ is dual to the model with $-B$
under the transformation (\ref{eq:dual2}) and the model is
self-dual in the space $B=0$, i.e.,
\begin{equation}\label{eq:self-dual3}
U-V_{\perp}-t_{\rm pair}-(2J_{\perp}-J_{\perp}^z)/4=0.
\end{equation}
The transformed coupling parameters are given by
\begin{eqnarray}
\bar{U} &=& \frac{1}{2}(U + V_{\perp} + t_{\rm pair}) +
\frac{1}{8}(2J_{\perp}-J_{\perp}^z),
\nonumber \\
\bar{V}_\perp &=& \frac{1}{4}(U + 3 V_{\perp} - t_{\rm pair}) -
\frac{1}{16}(2J_{\perp}-J_{\perp}^z),
\nonumber \\
\bar{J}_\perp &=& U - V_{\perp} - t_{\rm pair} +
\frac{1}{4}(2J_{\perp}+J_{\perp}^z),
\\
\bar{J}_\perp^z &=& - U + V_{\perp} + t_{\rm pair} +
\frac{1}{4}(2J_{\perp}+3J_{\perp}^z),
\nonumber \\
\bar{t}_{\rm pair} &=& \frac{1}{2}(U - V_{\perp} + t_{\rm pair}) -
\frac{1}{8}(2J_{\perp}-J_{\perp}^z). \nonumber
\end{eqnarray}
This duality relation leads to the conclusion that, if CDW or DDW
order, for example, exists in a certain parameter region, spin
current (DSDW) or SDW order exists in a dual region, respectively.
Note that even if we start from a spin isotropic model the dual
model is spin anisotropic and hence spin symmetry breaking can
occur in the dual model. The phase diagram must be symmetric with
the self-dual space (\ref{eq:self-dual3}) and the transitions
between dual phases, if ever, locate exactly on the self-dual
space.

The spin-charge duality relation associated with the
transformation (\ref{eq:dual_comb}) can be obtained from the
combination of two transformations discussed above. Thereby one
can arrive at a direct duality relation between orders in the
charge and spin sectors. The self-dual space is the intersection
of Eqs.\ (\ref{eq:self-dual}) and (\ref{eq:self-dual3}), i.e., $U
- V_{\perp} + \frac{1}{4}J_{\perp}^z = 0$ and
$\frac{1}{2}J_{\perp} + t_{\rm pair} = 0$.

In the self-dual spaces (\ref{eq:self-dual}) and
(\ref{eq:self-dual3}), the model Hamiltonian is invariant under
the continuous transformations with the unitary operators
$U_I(\theta)$ and $U_{II}(\theta)$, respectively, and hence U(1)
symmetric. [This can be easily seen in Eq.\
(\ref{eq:H_coupling})]. Because of the U(1) symmetry, a rigorous
theorem\cite{Momoi} concludes that the dual orders that are
continuously transformed to each other disappear on the self-dual
models in one dimension, in general\cite{note2}, and the phase
transition between the dual phases, if ever, is of second order.
The $A$- and $B$-terms in the interaction (\ref{eq:H_coupling})
serve as symmetry-breaking perturbations for the U(1) symmetry of
$U_I(\theta)$ and $U_{II}(\theta)$, respectively. When a
symmetry-breaking perturbation is relevant, it induces order and a
gap determining the criticality of the phase transition.

%In bosonization arguments, the models at critical points are governed
%by a U(1) symmetric fixed point.??? In the present model, however,
%the bare Hamiltonian already has the exact U(1) symmetry.

%discussion
In summary, we have developed duality transformations for electron
systems, which lead to duality relations between various exotic
phases shown in Fig.\ \ref{fig:duality}. These duality relations
have an analogy with the spin-chirality duality transformation we
and co-workers introduced for the spin
ladder\cite{HikiharaMH,MomoiHNH}, but are applicable to a much
wider variety of systems. The transformations clarify for the
generalized two-leg Hubbard ladder that the stability of
unconventional density-wave orders such as staggered flux (DDW)
and circulating spin current (DSDW) is equal to that of
conventional density-wave orders in the dual spaces. Recent
large-scale numerical analyses\cite{VojtaHN,Schollwoeck} reported
the appearance of CDW and DDW phases\cite{Note} under doping.
Application of the duality relations to these results immediately
concludes that SDW and DSDW phases stably exist under doping in
dual parameter spaces. Further arguments and possible applications
of the duality transformations will be reported elsewhere.

%One can apply the duality transformations to other various
%states as well. For example, the duality transformation (I)
%changes $s$-wave pairing superconductivity to $d$-wave one, and a
%spin liquid Mott insulator, "S-Mott" state, to "D-Mott"
%state\cite{Note0}, whereas the duality (II) does not change these
%states. Further arguments and applications will be published
%elsewhere.

Finally we stress that the duality transformations
(\ref{eq:dual1}) and (\ref{eq:dual2}) can be applied to a variety
of systems in which bond degrees of freedom are important, since
they are based only on simple gauge transformations of bond
operators. Applications to double-layer systems, and two- and
three-dimensional systems with orbital degeneracy are of interest.

The authors would like to thank H.\ Tsunetsugu, M.\ Tsuchiizu, A.\
Furusaki, and Y.\ Ohashi for stimulating discussions. They are
also grateful to M.\ Nakamura for useful comments on the unitary
representation of the duality transformations. T.M. was supported
in part by Monkasho (MEXT) in Japan through Grand No.\ 14540362.

%\vspace*{-3mm}


\begin{thebibliography}{99}
\bibitem{note2} An exception occurs only if susceptibility of the
generator for the rotation is diverging. (See Ref.\ \cite{Momoi}.)
In that case, the orders can survive in the self-dual models,
though there are gapless excitations at the transition points
associated with continuous symmetry breaking.
\bibitem{Note} Note that these two phases reported in Refs.\
\cite{VojtaHN} and \cite{Schollwoeck} are not dual to each other.
The CDW state in Ref.\ \cite{VojtaHN} has an oscillation with
fixed wave vector $q=\pi$ for all fillings, but the DDW state in
Ref.\ \cite{Schollwoeck} shows incommensurate oscillation under
doping. The model discussed in Ref.\ \cite{VojtaHN} contains
inter-rung nearest neighbor Coulomb repulsion $V_\parallel$.
\end{thebibliography}
\end{document}